\documentclass[aps,preprint,prd,showpacs,nofootinbib]{revtex4}

\usepackage{latexsym}
\usepackage{amsmath}
\usepackage{graphicx}
\usepackage{subfigure}
\usepackage{dcolumn}
\usepackage{bm}
\usepackage{amssymb}
\usepackage{latexsym}
\usepackage{ulem}
\usepackage{color}
\usepackage[colorlinks,linkcolor=magenta,anchorcolor=blue,citecolor=blue]{hyperref}

\def\be{\begin{equation}}
\def\ee{\end{equation}}
\def\ba{\begin{eqnarray}}
\def\ea{\end{eqnarray}}

\def\nn{\nonumber}
\def\lf{\left}
\def\rt{\right}

\begin{document}

\title{ Is there an effect of a nontrivial $c_T$ during inflation? }

\author{Yong Cai$^{1}$\footnote{caiyong13@mails.ucas.ac.cn}}
\author{Yu-Tong Wang$^{1}$\footnote{wangyutong12@mails.ucas.ac.cn}}
\author{Yun-Song Piao$^{1,2}$\footnote{yspiao@ucas.ac.cn}}

\affiliation{$^1$ School of Physics, University of Chinese Academy of
Sciences, Beijing 100049, China}

\affiliation{$^2$ State Key Laboratory of Theoretical Physics, Institute of Theoretical Physics, \\
Chinese Academy of Sciences, P.O. Box 2735, Beijing 100190, China}

\begin{abstract}

Recently, we have shown that the propagation speed $c_T$ of
the primordial gravitational waves (GWs) might be nontrivially
varying during inflation, which could induce local oscillations
in the power spectrum of primordial GWs. In this paper, we
numerically confirm that, although with a disformal
redefinition of the metric the nontrivial $c_T$ may be set as
unity, the power spectrum in the frame with $c_T=1$ is completely
the same as that in the original disformal frame, i.e., the
oscillating shape in the power spectrum is still reserved, since
here the effect of $c_T$ is actually encoded in the
nontrivially varying Hubble parameter. In addition, we also
clarify how to obtain a blue-tilted GWs spectrum by imposing a
rapidly decreasing $c_T$ during inflation.

\end{abstract}
\maketitle

\section{Introduction}

Inflation \cite{Guth:1980zm} has been accepted as the theory of
the early Universe by most cosmologists for its success in
resolving several classical problems in the standard hot big bang
cosmology and predicting the primordial scalar and tensor
perturbations. In the recent years, high-precision observations of
the cosmic microwave background (CMB) strongly support the
presence of a nearly scale-invariant primordial scalar
perturbation. However, the primordial tensor perturbations,
i.e., the primordial gravitational waves (GWs) \cite{Gri:1974},
still need to be confirmed after the analysis of the most resent
polarization results of BICEP2, Keck Array, and Planck
\cite{Ade:2015tva}\cite{Ade:2015lrj}.

The detection of the primordial GWs is of great significance for
verifying general relativity (GR) and strengthening our confidence in
inflation. More interestingly, the primordial GWs spectrum may
provide us with rich information about the theory of gravity at an
extremely high energy scale. Therefore, a lot of experiments aimed
at searching for primordial GWs is scheduled in the coming
decades. With the experience of Planck and BICEP2, the CMB B-mode
detections might be promising to bring us some news about the
primordial GWs in the coming years. Meanwhile, the planning space-based
GWs detectors will also be able to place constraints on the
middle-frequency GWs background.

In GR, the propagating speed $c_T$ of GWs is the same as the speed of
light, and thus can naturally be set as unity. However, in modified
gravity theories, such as Horndeski theory
\cite{Horndeski:1974wa}\cite{Deffayet:2011gz}\cite{Kobayashi:2011nu},
beyond Horndeski theories
\cite{Gleyzes:2014dya},\cite{Gao:2014soa}, Gauss-Bonnet gravity
\cite{Nojiri:2005jg}\cite{Satoh:2008ck}\cite{Oikonomou:2015qha},
next-to-leading-order fluid theory \cite{Ballesteros:2014sxa}, and
the low-energy effective string theory with higher-order
corrections
\cite{Met:1987}\cite{Antoniadis:1993jc}\cite{Kawai:1998ab}\cite{Cartier:1999vk},
see also \cite{Piao:2003hh}\cite{Maeda:2004vm}, $c_T$ may deviate
from unity. Recently, we have shown in
\cite{Cai:2015ipa}\cite{Cai:2015dta} that since $c_T$ and the
Hubble radius($\sim H^{-1}$) determine the horizon-crossing time
of the primordial perturbation modes, the nontrivial variation of
$c_T$ could induce local oscillations in the power spectrum of
primordial GWs, which can leave some observable imprints in the
CMB B-mode spectrum, e.g. the oscillations around the
recombination peak, see also
\cite{Amendola:2014wma}\cite{Raveri:2014eea}\cite{Jimenez:2015bwa}
for the effect of modified gravity at late time.

Recently, it has been argued in \cite{Creminelli:2014wna} that
with a disformal redefinition of the metric, $c_T$ always
can be set as unity and the Planck mass can be time independent, which
makes the GWs spectrum formally the same as that in GR, see also
\cite{Watanabe:2015uqa}\cite{Baumann:2015xxa}. In the new
frame with $c_T=1$, the primordial GWs spectrum is $P_T=2{\tilde
H}^2/\pi^2M_p^2$. Since the power spectrum of primordial GWs is
independent of the choice of frames, e.g.,
\cite{Minamitsuji:2014waa}\cite{Tsujikawa:2014uza}\cite{Li:2015hga},
one might think that there should be no observable effect induced
by $c_T$ in the power spectrum of primordial GWs.

In this paper, we show that, although the propagation speed $c_T$
of the primordial GWs could be set as unity with a proper
disformal rescaling, it does not mean that there are no observable
effects induced by $c_T$. We confirm numerically that the power
spectrum in the frame with $c_T=1$ is completely the same as that in
the original disformal frame, i.e., the oscillating shape in the
power spectrum is still reserved, since the effect of $c_T$ has
been encoded in the nontrivially varying Hubble parameter $\tilde
H$ in the new frame. We also suggest an inflationary model, in
which the scalar spectrum is scale invariant with a slight red
tilt while the local spectrum of GWs may be slightly or strongly
blue without violating the null energy condition, by imposing a
rapidly decreasing $c_T$.

\section{Primordial GWs in the disformal frame}

We follow the effective field theory of inflation
\cite{Cheung:2007st}, and in unitary gauge the action is
\ba S & = & \int d^4x \sqrt{-g} \Big[{M_p^2\over 2}R-c_1(t)-c_2(t)g^{00}\nn\\
& &\,\,\,\,\,\,\,\,\,\,\,\,\,\,\,\,\,\,\,\,\,\,\,\,\,\,\,\,\,- \,
{M_p^2\over 2}\left(1-{1\over c_T^{2}(t)}\right)\left(\delta
K_{\mu\nu}\delta K^{\mu\nu}-\delta K^2 \right)  \Big]\,,
\label{action}\ea where the parameters $c_1$ and $c_2$ are set by
the Friedmann equations, which are the functions of the Hubble
parameter $H$ and ${\dot H}$; $K_{\mu\nu}$ is the extrinsic
curvature of the spatial slices; the first line describes the
slow-roll inflation with the propagation speed of primordial GWs
$c_T=1$; and the second line modifies it with $c_T(t)$, but does not affect the
scalar perturbation, as was pointed out in
\cite{Creminelli:2014wna}. The tensor mode $\gamma_{ij}$ satisfies
$\gamma_{ii}=0$, and $\partial_i \gamma_{ij}=0$, and its quadratic
action for (\ref{action}) is
\be S^{(2)}=\int d\tau d^3x
{M_p^2a^2\over 8}\lf[{1\over c_T^2}\lf(\frac{d\gamma_{ij}
}{d\tau}\rt)^2-(\vec{\nabla}\gamma_{ij})^2\rt]\,,
\ee
where
$\tau=\int dt/a$.

In this disformal frame, the power spectrum of primordial
GWs is calculated as follows. The Fourier series of $\gamma_{ij}$
is \be \gamma_{ij}(\tau,\mathbf{x})=\int \frac{d^3k}{(2\pi)^{3}
}e^{-i\mathbf{k}\cdot \mathbf{x}} \sum_{\lambda=+,\times}
\hat{\gamma}_{\lambda}(\tau,\mathbf{k})
\epsilon^{(\lambda)}_{ij}(\mathbf{k}), \ee where $
\hat{\gamma}_{\lambda}(\tau,\mathbf{k})=
\gamma_{\lambda}(\tau,k)a_{\lambda}(\mathbf{k})
+\gamma_{\lambda}^*(\tau,-k)a_{\lambda}^{\dag}(-\mathbf{k})$, the
polarization tensors $\epsilon_{ij}^{(\lambda)}(\mathbf{k})$
satisfy $k_{j}\epsilon_{ij}^{(\lambda)}(\mathbf{k})=0$,
$\epsilon_{ii}^{(\lambda)}(\mathbf{k})=0$,
$\epsilon_{ij}^{(\lambda)}(\mathbf{k})
\epsilon_{ij}^{*(\lambda^{\prime}) }(\mathbf{k})=\delta_{\lambda
\lambda^{\prime} }$ and $\epsilon_{ij}^{*(\lambda)
}(\mathbf{k})=\epsilon_{ij}^{(\lambda) }(-\mathbf{k})$; the
commutation relation of the annihilation and creation operators
$a_{\lambda}(\mathbf{k})$ and
$a^{\dag}_{\lambda}(\mathbf{k}^{\prime})$ is $[
a_{\lambda}(\mathbf{k}),a_{\lambda^{\prime}}^{\dag}(\mathbf{k}^{\prime})
]=\delta_{\lambda\lambda^{\prime}}\delta^{(3)}(\mathbf{k}-\mathbf{k}^{\prime})$.
Thus, the equation of motion for $u(\tau,k)$ is given as \be
\frac{d^2u}{d\tau^2}+\left(c_T^2k^2-\frac{d^2z_T/d\tau^2}{z_T}
\right)u=0, \label{eom1} \ee where \be {u}(\tau,k)=
\gamma_{\lambda}(\tau,k) {z_T}, \quad z_T= {aM_p { c_T^{-1} }\over
2}.\label{zt} \ee Initially, the perturbations are deep inside the
horizon, i.e., $c_T^2k^2 \gg \frac{d^2z_T/d\tau^2}{z_T}$. The
initial condition is assumed to be the Bunch-Davies vacuum, thus,
$u\sim \frac{1}{\sqrt{2c_T k} }e^{-i c_T k\tau}$. The power
spectrum of primordial GWs is \be
P_T=\frac{k^3}{2\pi^2}\sum_{\lambda=+,\times} \lf|\gamma_{\lambda}
\rt|^2=\frac{4k^3}{\pi^2 M_p^2}\cdot\frac{c_T^2}{ a^2} \lf|u
\rt|^2, \quad aH/(c_Tk) \gg 1.\label{pt} \ee   Here, $c_T$ is
not required to be a constant. The effect of varying sound speed
$c_S$ of scalar perturbation on the scalar power spectrum has been
widely investigated (see e.g., \cite{Nakashima:2010sa},
\cite{Park:2012rh} and \cite{Achucarro:2014msa}).

When $c_T$ varies drastically with time, Eq.(\ref{eom1}) can hardly  be
solved analytically. Therefore, as in
Ref.\cite{Cai:2015ipa}, we will exhibit the numerical results
below. We introduce the variable $\alpha=\ln a$ and write
Eq.(\ref{eom1}) as \be u_{\alpha\alpha}+\lf(1+\frac{ H_{\alpha}
}{H} \rt)u_{\alpha}+\frac{1}{a^2 H^2
}\lf(c_T^2k^2-\frac{d^2z_T/d\tau^2}{z_T} \rt)u=0,\label{eomold}
\ee where \be \frac{d^2z_T/d\tau^2}{z_T}=a^2H^2\Big[2+\frac{
H_{\alpha} }{H}-\lf(3+\frac{ H_{\alpha} }{H}\rt)\frac{c_{T\alpha}
}{c_T}+2\lf(\frac{c_{T\alpha} }{c_T}\rt)^2-\frac{c_{T\alpha\alpha}
}{c_T}  \Big], \ee with the initial condition \be u|_{\alpha \ll
\ln(\frac{k}{H})}=\frac{1}{\sqrt{2k} }, \quad \frac{\partial
u}{\partial \alpha}\Big|_{\alpha \ll \ln(\frac{k}{H})}=-i
\sqrt{\frac{k}{2} }\frac{1}{e^{\alpha} H}\Big|_{\alpha \ll
\ln(\frac{k}{H})},\label{initial1} \ee where the subscript
`$\alpha$' denotes $\partial/\partial \alpha$. For simplicity,
we set $c_T=1$ initially and assume that the background of
slow-roll inflation is described by constant $\epsilon$, and the
Hubble parameter is $H(\alpha)/M_p=A_H\cdot e^{-\epsilon \cdot
\alpha}$, in which $A_H$ is constant.

The behavior of $c_T$ is required for solving the perturbation
equation (\ref{eomold}) numerically. In \cite{Cai:2015ipa},
it was found that in a certain setup the leading-order
$\alpha^\prime$ correction $\sim e^{-\phi}R^2_{GB}$ in heterotic
string theory may lead to a short-time but drastic variation of
$c_T$. However, to clearly clarify the nontrivial effect
of $c_T$ on the power spectrum of primordial GWs, here, we will not
get involved in the specific models. Instead, inspired by
\cite{Cai:2015ipa}, we will focus on two phenomenological
templates as follows.

\subsection{Oscillating $c_T$ with $c_T=1+\frac{c_{T*}}{1+(\alpha-\alpha_*)^n}\cdot
\sin[B(\alpha-\alpha_*)]$ and the oscillating spectrum}

In this template, $c_{T*}$ and $B$ depict the amplitude and
frequency of local oscillation of $c_T$, respectively, and $\alpha_*$
determines the time when the drastic variation happens, i.e., the
scale where the nontrivial effect of $c_T$ appears in the GWs
spectrum.

\begin{figure}[htbp]
\subfigure[~~$c_T(\alpha)$]{\includegraphics[width=.45\textwidth]{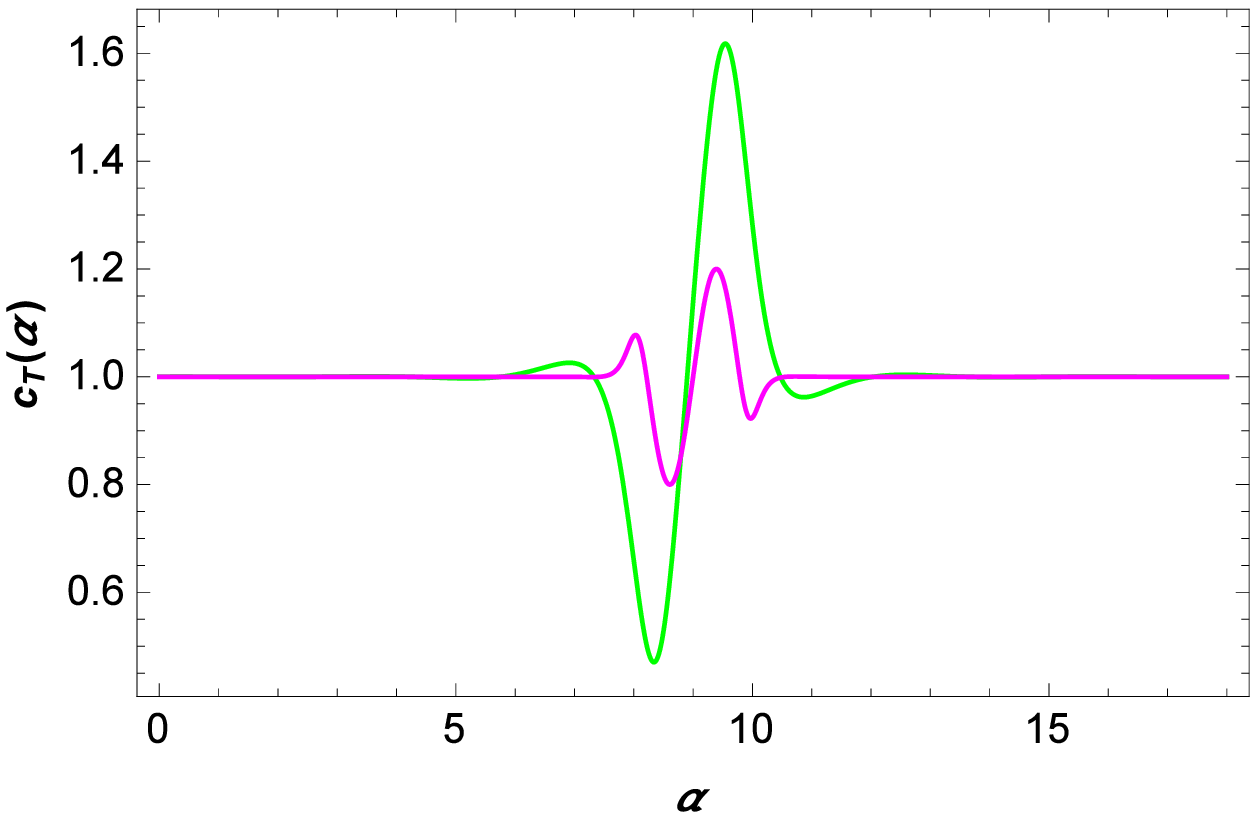}
} \subfigure[~~The power spectrum in the disformal frame.
]{\includegraphics[width=.5\textwidth]{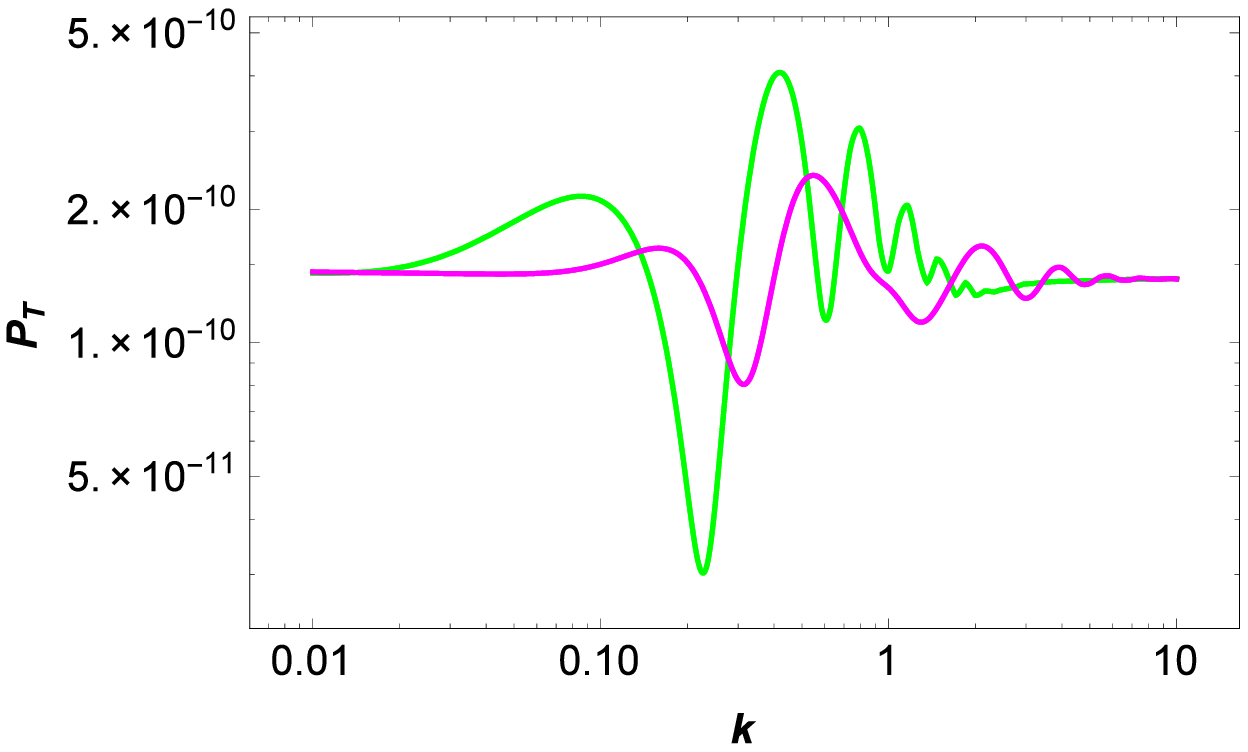} }
\caption{$c_T(\alpha)=1+\frac{c_{T*}}{1+(\alpha-\alpha_*)^n}
\sin[B(\alpha-\alpha_*)]$, where $\alpha_*=9$, $\epsilon=0.003$
and $A_H=2.72\times10^{-5}$. We set $c_{T*}=0.2$, $B=4$, $n=10$ for the
magenta lines and $c_{T*}=0.7$, $B=2$, $n=4$ for the green lines.} \label{fig01}
\end{figure}

In Fig.\ref{fig01}, we see that the local oscillation is present
in $P_T$, which is obviously induced by the nontrivial oscillation
of $c_T$. Actually, this result is anticipated, which may be
analytically showed as follows.

We assume that $c_T$ undergoes an ideally steplike jump (i.e., initially, the speed of GWs is a constant $c_{T1}$, then, at sometime it suddenly jumps to another constant value $c_{T2}$). In such a case, the power spectrum
can be given as \cite{Cai:2015dta}
\be
P_T=P_T^{inf}\cdot\frac{f(k,k_0,x)}{c_{T1}^3 }, \ee where \be
f(k,k_0,x)\approx
\frac{1}{x^2}\lf[1+(\frac{1}{x^2}-1)\cos^2(\frac{k}{k_0})\rt] \ee
for $k> k_0$, $k_0$ is the wave number of the GWs mode
exiting the horizon when $c_T$ jumps, $x=c_{T2}/c_{T1}$, and
$P_T^{inf}=2H^2_{inf}/\pi^2M_p^2$ is the standard result of
slow-roll inflation with $c_T=1$. Obviously, $f(k,k_0,x)$ will
oscillate between $1/x^2$ and $1/x^4$. Hence, for the more general
case in which $c_T$ experiences drastic variation, we roughly have \be
\lf(\frac{P_T}{P_T^{inf}}\rt)_{max}\approx\frac{
c_{T\,ini}}{c^4_{T\,min}},\qquad
\lf(\frac{P_T}{P_T^{inf}}\rt)_{min}\approx\frac{
c_{T\,ini}}{c^4_{T\,max}}, \ee namely, $P_T$ oscillates between
$P_T^{inf}\cdot c_{T\,ini}/c^4_{T\,max}$ and $P_T^{inf}\cdot
c_{T\,ini}/c^4_{T\,min}$, where ``max" and ``min" denote the maximal
and minimal values of corresponding quantities during oscillating,
respectively, and ``ini" denotes the initial value. In
Fig.\ref{fig01}, for the green line,
$c_{T\,ini}/c^4_{T\,min}\simeq 1/(0.5)^4\sim 10$ and
$c_{T\,ini}/c^4_{T\,max}\simeq 1/(1.6)^4\sim 0.1$, and thus the
oscillating amplitude of $P_T$ obeys $0.1\lesssim
P_T/P_T^{inf}\lesssim 10$, which is consistent with the
corresponding power spectrum.

The similar oscillating shape in the primordial GWs spectrum
has been obtained in a stringy model \cite{Cai:2015ipa}, in which
the variation of $c_T$ similar to that in Fig.\ref{fig01} was
implemented. There are also other mechanisms, which may endow the
power spectrum of GWs produced during inflation with features,
such as the particle production
\cite{Cook:2011sp}\cite{Senatore:2011sp}, and the pseudoscalar field
interacting with gauge fields
\cite{Sorbo:2011ja}\cite{Mukohyama:2014gba}\cite{Ferreira:2014zia}\cite{Namba:2015gja}.
However, their signals are entirely different from the oscillation
shown here. In addition, some theories of modified gravity can
also alter the primordial GWs spectrum,
e.g., \cite{Johnson:2015tfa}\cite{Fasiello:2015csa}.

\subsection{Dipping $c_T$ with $c_T=1-c_{T*}e^{-D (\alpha-\alpha_*)^2}$ and the blue spectrum}

In this template, $c_{T*}$ and $D$ depict the depth and width of the
dipping, respectively, and similar to that in II.A, $\alpha_*$
determines the time when the dipping happens.

\begin{figure}[htbp]
\subfigure[~~$c_T(\alpha)$]{\includegraphics[width=.45\textwidth]{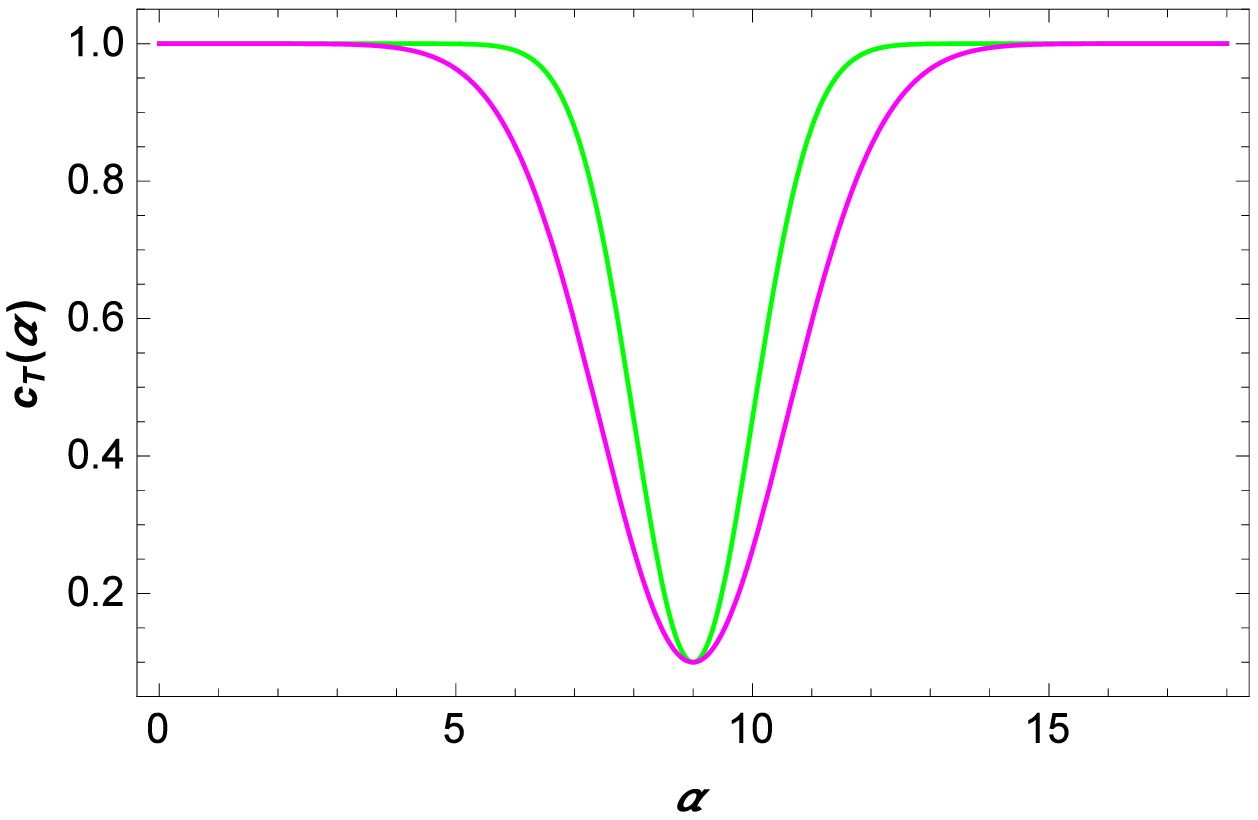}
} \subfigure[~~The power spectrum in the disformal frame.
]{\includegraphics[width=.5\textwidth]{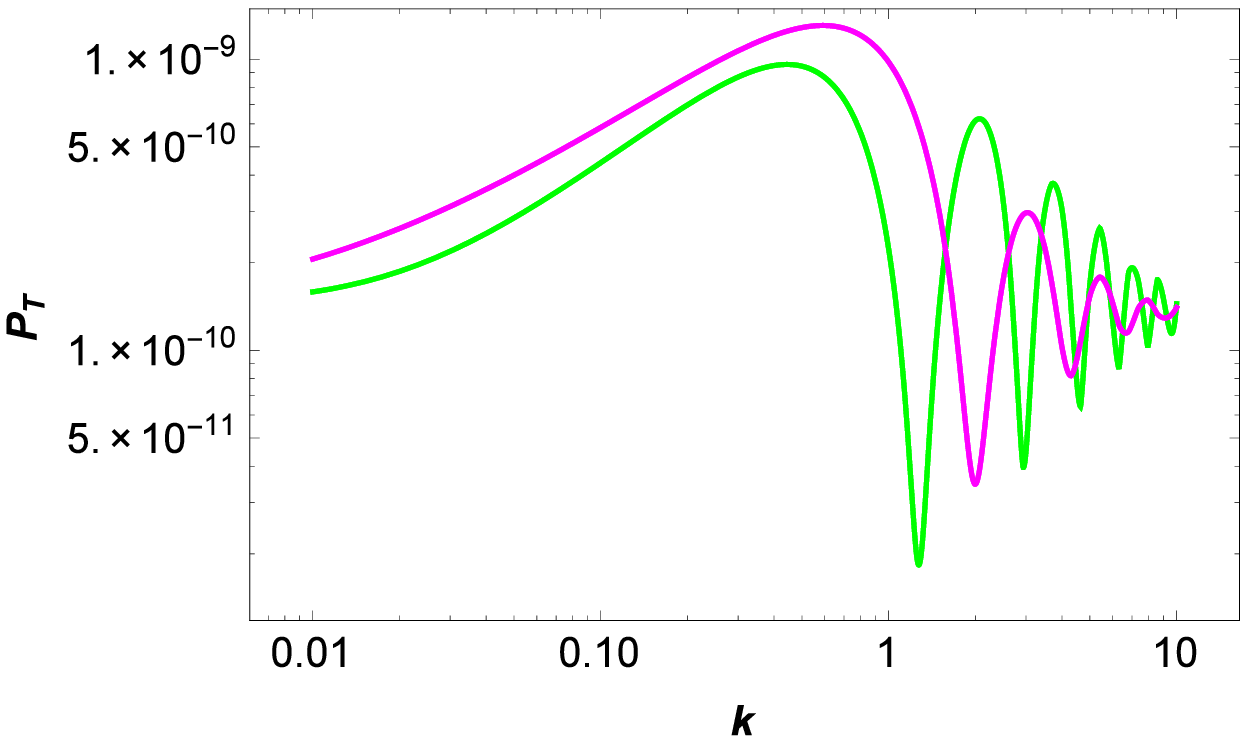}
} \caption{$c_T(\alpha)=1-c_{T*}e^{-D (\alpha-\alpha_*)^2}$, where
$c_{T*}=0.9$, $\alpha_*=9$, $\epsilon=0.003$ and
$A_H=2.72\times10^{-5}$. We set $D=0.2$ for the magenta lines and
$D=0.5$ for the green lines.} \label{fig02}
\end{figure}

In Fig.\ref{fig02}, we see that the local bump is present in
$P_T$, i.e., rapidly increases and then returns with an
oscillating modulation. The rapid increasing of $P_T$ implies the
spectrum is blue tilted at corresponding region. This tilt is
induced by the rapid decrease of $c_T$ during dipping, which may
be estimated as follows.

We, for simplicity, assume that locally $c_T$ decrease like \be c_T=
(-H_{inf}\tau)^p, \label{cTn}\ee in which $p>0$. It may be
rewritten as ${{\dot c}_T\over H_{inf}c_T}=-p$.

We set $dy=c_Td\tau$ \cite{Khoury:2008wj}, and thus the equation of
motion of $u(y,k)$ is \be u_{yy}+\left(k^2-\frac{z_{T,yy}}{z_T}
\right)u=0, \label{eom4} \ee where ${u}(y,k)=
\gamma_{\lambda}(y,k) {z_T}$ and $z_T= {aM_p { c_T^{-1/2} }\over
2}$. On superhorizon scales, its solution is \be |u|^2\simeq {1\over
2k}(-ky)^{1-{3+2p\over 1+p}}. \ee Thus the power spectrum of GWs is
\be P_T={4k^3\over \pi^2 M_P^2} {c_T|u^2|\over
a^2}={2H_{inf}^2{ (1+p)^2}\over \pi^2 c_T M_P^2} (-ky)^{p\over 1+p}, \ee
where $y={c_T\over { 1+p} }\tau={-}{c_T\over { (1+p)}aH_{inf} }$. Thus, \be n_T={p\over 1+p}
\label{nTB}\ee is blue tilted. Here, we have neglected the running
of $H_{inf}$;  otherwise, it will contribute $-2\epsilon$. When $p\ll 1$,
i.e., $c_T$ is slowly varying, the result is consistent with that
in Ref.\cite{DeFelice:2014bma}. However, ours is valid not only
for $p\ll 1$ but also for an arbitrary positive value of $p$. In
addition, it should be mentioned that $0<n_T\leqslant 1$, i.e.,
$n_T$ cannot be larger than 1.

The blue GWs spectrum is interesting, since it boosts the
possibility that the primordial GWs background is
detectable \cite{Zhao:2013bba}\cite{Boyle:2007zx}, see also \cite{Brustein:1995ah}\cite{Gasperini:2002bn} for the discussions in the pre-big bang scenario. How to
obtain a blue GWs spectrum without the ghost instability
while reserving a scale-invariant scalar spectrum with a red tilt is
still a significant issue, e.g.,
\cite{Piao:2003ty}\cite{Creminelli:2006xe}\cite{Wang:2014kqa}\cite{Cai:2014uka},
and also \cite{Cannone:2014uqa}. Here, we actually suggest such an
alternative. The background is given by the first line in
(\ref{action}), which is the slow-roll inflation. The scalar
perturbation is not affected by the decrease of $c_T(t)$. Thus its
spectrum may be slightly red tilted, as required by the
observations. However, the spectrum of GWs, which is given by (\ref{nTB}),  may be slightly
or strongly blue. The more rapidly
$c_T$ decreases, the bluer the spectrum is. We will discuss this
scenario again in the end of Sec. III.

\section{Primordial GWs in the Einstein frame}

In Ref.\cite{Creminelli:2014wna}, it is demonstrated for
(\ref{action}) (and also those cases that the quadratic operators has at
most two-spatial derivatives) that the propagation speed
$c_T^2(t)$ of GWs can be set to unity with a disformal redefinition
of metric \be g_{\mu\nu}\rightarrow
c_T^{-1}\left[g_{\mu\nu}+{(1-c_T^2)}n_{\mu}n_{\nu}\right]\,.
\label{gmunu1} \ee
Then, the line element is $ds^2=-c_Tdt^2+c_T^{-1}a^2d\mathbf{x}^2$.
It is convenient to define \be \tilde{t}\equiv\int
c_T^{1/2}dt,\qquad \tilde{a}(\tilde{t})\equiv c_T^{-1/2}a(t), \ee
which makes the line element be $ds^2=-d{\tilde t}^2+{\tilde
a}^2d{\mathbf{x}}^2$. In the new frame, i.e., the
Einstein frame, the quadratic action of $\gamma_{ij}$ becomes (see,
e.g., \cite{Tsujikawa:2014uza}\cite{Baumann:2015xxa}) \be
S^{(2)}=\int d\tilde{\tau} d^3x {M_p^2\tilde{a}^2\over
8}\lf[\lf(\frac{d\gamma_{ij}
}{d\tilde{\tau}}\rt)^{\,2}-(\vec{\nabla}\gamma_{ij})^2\rt]\,,\label{newaction}
\ee where $d\tilde{\tau}=d\tilde{t}/\tilde{a}$.
Thus, we have $c_T=1$ in this frame. We can
calculate the power spectrum as follows.

The equation of motion for $u(\tilde{\tau},k)$ is \be
\frac{d^2u}{d\tilde{\tau}^2
}+\left(k^2-\frac{d^2\tilde{z}_T/d\tilde{\tau}^2}{\tilde{z}_T}
\right)u=0,\label{eom2} \ee where \be {u}(\tilde{\tau},k)=
\gamma_{\lambda}(\tilde{\tau},k)\cdot {\tilde{z}_T}, \quad
\tilde{z}_T= {\tilde{a}M_p\over 2}. \ee The initial condition is
still the Bunch-Davies vacuum $u\sim\frac{1}{\sqrt{2k}
}e^{-ik\tilde{\tau}}$. The power spectrum of primordial GWs can be
written as \be P_T=\frac{k^3}{2\pi^2}\sum_{\lambda=+,\times}
\lf|\gamma_{\lambda} \rt|^2=\frac{4k^3}{\pi^2
M_p^2\tilde{a}^2}\cdot\lf|u \rt|^2, \quad \tilde{a}\tilde{H}/k \gg
1.\label{newPT} \ee When $\tilde{\epsilon
}=-\frac{d\tilde{H}/d\tilde{t} }{\tilde{H}^2}\ll 1$, we have \be
P_T=2{\tilde H}^2/\pi^2 M_P^2, \label{PT3}\ee where \be
\tilde{H}=\frac{d\tilde{a}/d\tilde{t} }{\tilde{a}
}=c_T^{-1/2}\lf(H_{inf}-\frac{dc_T/dt}{2c_T}\rt), \label{HH}\ee
which is just the standard result of slow-roll inflation but with
$\tilde H$ replacing $H_{inf}$. This result looks like a
conflict with that calculated in the disformal frame, in which the
drastic variation of $c_T(t)$ results in the nontrivial effect on
$P_T$, since the power spectrum of the primordial perturbations
should be independent of the choice of frames.

However, $\tilde{\epsilon }\ll 1$ implicitly assumes
$\frac{\dot{c_T}}{Hc_T}\ll1$. When $c_T$ experiences some drastic
variations, the condition $\tilde{\epsilon }\ll1$ is usually
broken, and thus (\ref{PT3}) cannot be applied to such cases.
Actually, since the the evolution of $c_T$ in the disformal frame
is totally encoded in $\tilde{H}$, see (\ref{HH}), a slow-roll
inflation with $c_T(t)$ in the disformal frame may no longer be
the slow-roll model in the new frame with $c_T\equiv1$.
Below, we will numerically confirm that the results in both frames
are equivalent.

Again, we write Eq.(\ref{eom2}) as  \be
u_{\tilde{\alpha}\tilde{\alpha}}+\lf(1+\frac{
\tilde{H}_{\tilde{\alpha}} }{\tilde{H}}
\rt)u_{\tilde{\alpha}}+\lf(\frac{k^2}{\tilde{a}^2 \tilde{H}^2
}-2-\frac{ \tilde{H}_{\tilde{\alpha}} }{\tilde{H}}
\rt)u=0,\label{eomnew} \ee with the initial conditions \be
u|_{\tilde{\alpha} \ll \ln(\frac{k}{\tilde{H}
})}=\frac{1}{\sqrt{2k} }, \quad \frac{\partial u}{\partial
\tilde{\alpha} }\Big|_{\tilde{\alpha} \ll \ln(\frac{k}{\tilde{H}
})}=-i \sqrt{\frac{k}{2} }\frac{1}{e^{\tilde{\alpha} } \tilde{H}
}\Big|_{\tilde{\alpha} \ll \ln(\frac{k}{\tilde{H}
})},\label{initial2} \ee which is actually equivalent to
(\ref{initial1}), since we have assumed $dc_T/d\tau=0$ initially,
where $\tilde{\alpha}=\ln \tilde{a}$ and the subscript
$\tilde{\alpha}$ denotes $\partial/\partial \tilde{\alpha}$.

\begin{figure}[htbp]
\subfigure[~~${\tilde H}$ and ${\tilde
\epsilon}$]{\includegraphics[width=.48\textwidth]{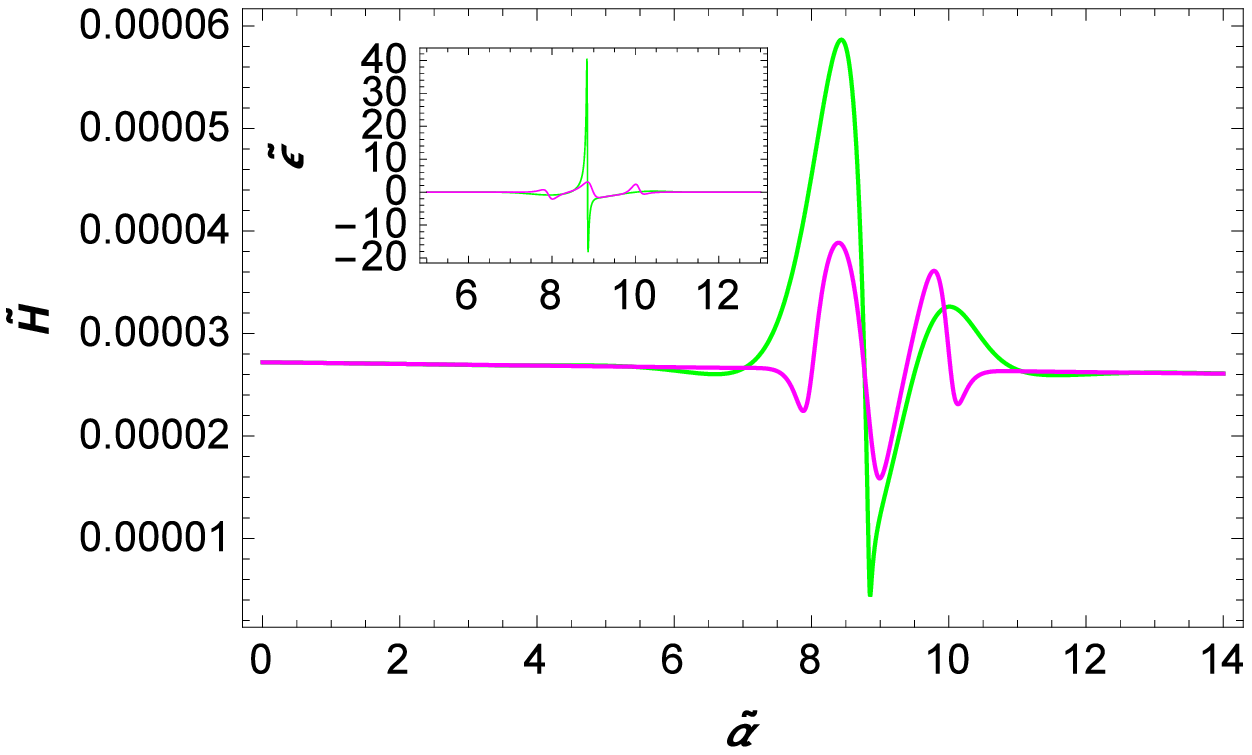}
} \subfigure[~~The power spectrum in the Einstein frame.
]{\includegraphics[width=.48\textwidth]{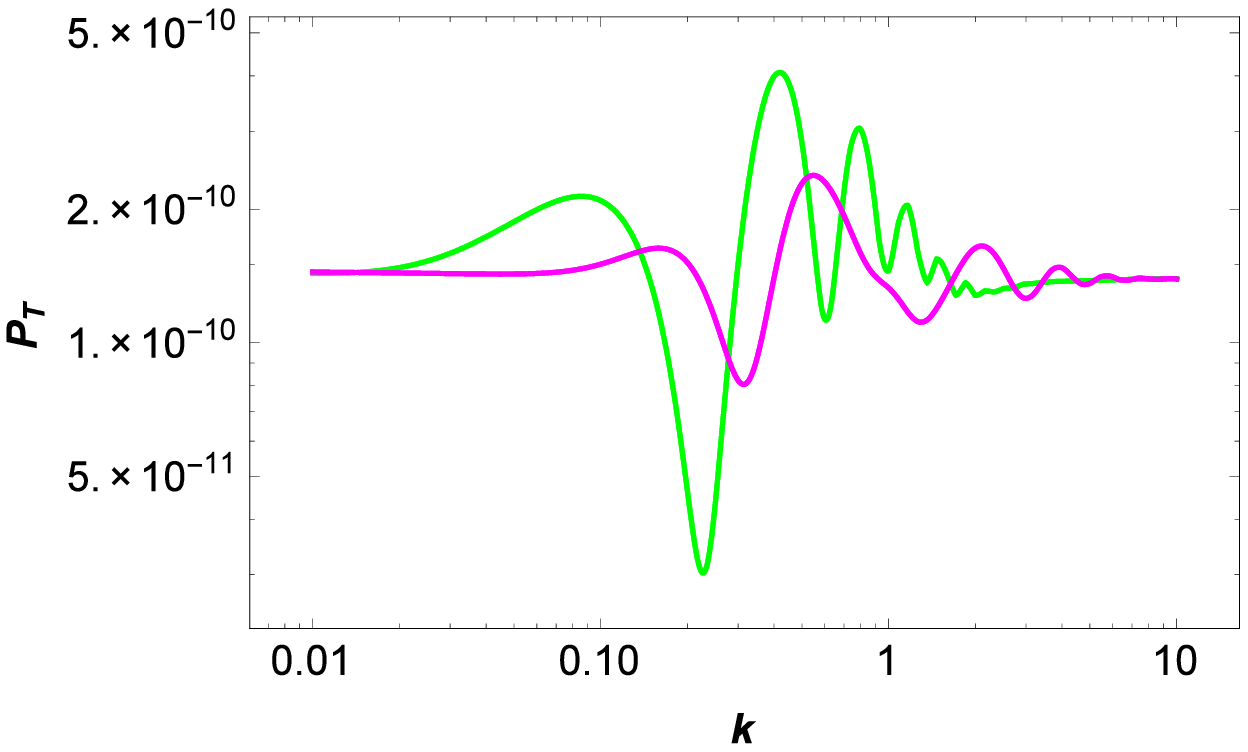} }
\caption{ $c_T(\alpha)=1+\frac{c_{T*}}{1+(\alpha-\alpha_*)^n}
\sin[B(\alpha-\alpha_*)]$, where $\alpha_*=9$, $\epsilon=0.003$
and $A_H=2.72\times10^{-5}$. We set $c_{T*}=0.2$, $B=4$, $n=10$ for the
magenta lines and $c_{T*}=0.7$, $B=2$, $n=4$ for the green lines.} \label{fig03}
\end{figure}

\begin{figure}[htbp]
\subfigure[~~${\tilde H}$ and ${\tilde
\epsilon}$]{\includegraphics[width=.48\textwidth]{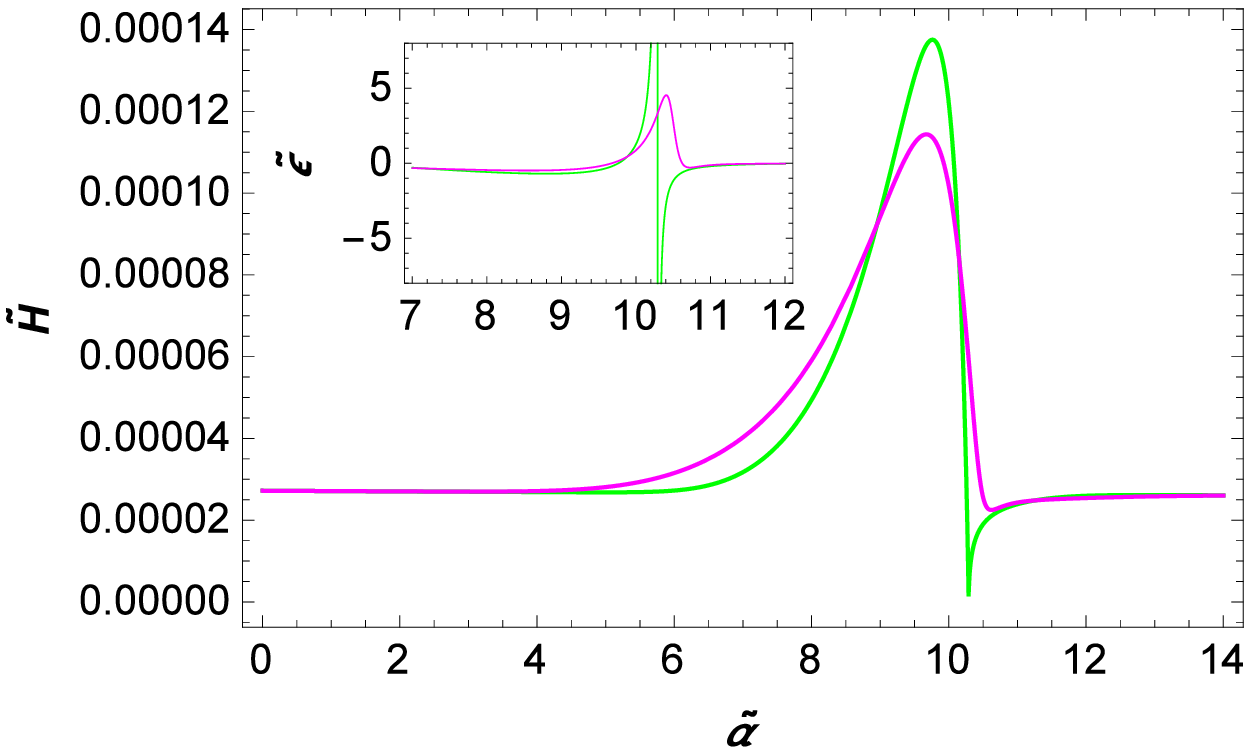}
} \subfigure[~~The power spectrum in the Einstein frame.
]{\includegraphics[width=.48\textwidth]{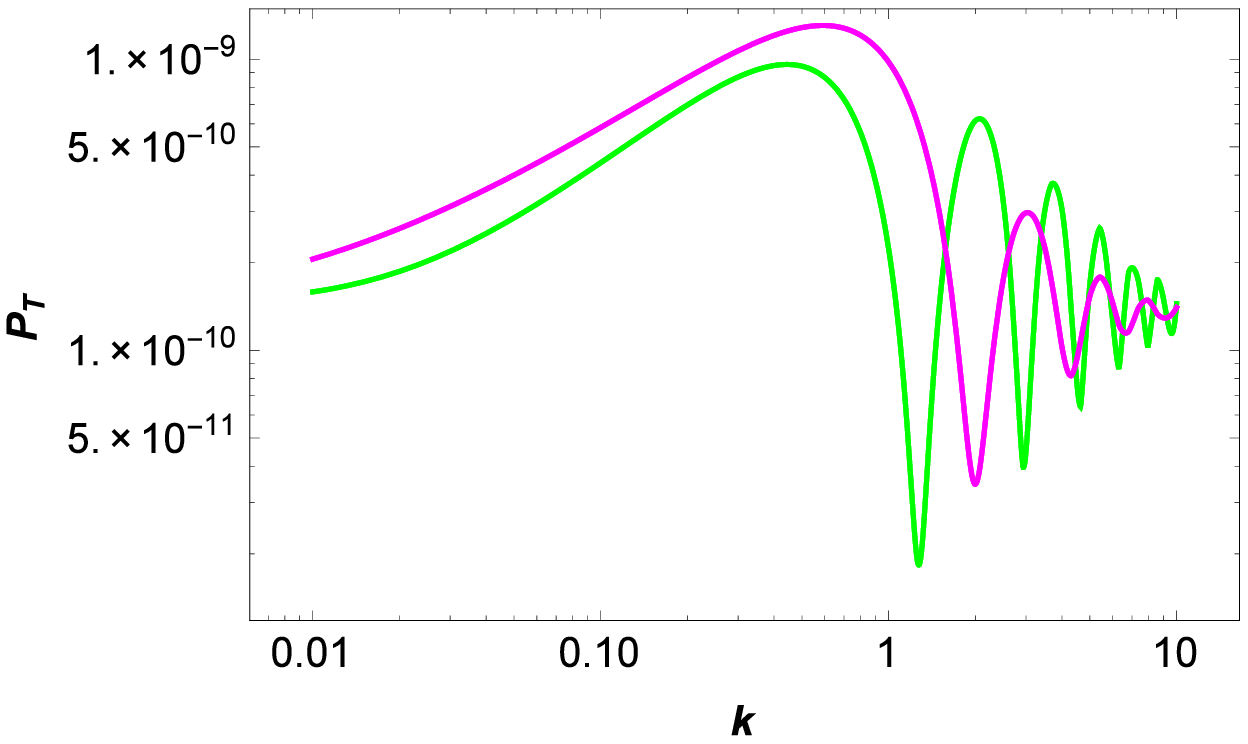}
} \caption{$c_T(\alpha)=1-c_{T*}e^{-D (\alpha-\alpha_*)^2}$, where
$c_{T*}=0.9$, $\alpha_*=9$, $\epsilon=0.003$ and
$A_H=2.72\times10^{-5}$. We set $D=0.2$ for the magenta lines and
$D=0.5$ for the green lines.} \label{fig04}
\end{figure}

Using the templates for $c_T$, we numerically solve
Eq.(\ref{eomnew}) and plot the evolutions of $\tilde{H}$,
$\tilde{\epsilon}$, and $P_T$ in Figs. \ref{fig03} and
\ref{fig04}. We see that the behavior of $c_T$ in the
disformal frame is fully encoded in $\tilde{H}(\tilde{\alpha})$,
which acts as the effective Hubble parameter in the new
frame, and $\tilde{\epsilon}=-\frac{d\tilde{H}/d\tilde{t}
}{\tilde{H}^2}\gg 1$ during some period though the slow-roll
parameter $\epsilon\ll 1$ in the original frame. This clearly indicates that, due to the
drastic variation of $c_T$, in the disformal frame the Universe is
in slow-roll inflation, but in the Einstein frame, it is not.
However, $P_T$ in both frames are completely same. Thus,
although the propagation speed $c_T$ of primordial GWs could be
set as unity, the effects of the nontrivial variation of $c_T$ on
the power spectrum of GWs is still reserved. Thus if we
see the oscillating shape in $P_T$, we
may attribute it to the dramatic evolution of $c_T$ in the disformal frame, or equally, to the dramatic evolution of
$\tilde{H}$ in the new frame.

We also numerically confirm that, although the sound speed of
scalar perturbation is altered as $\tilde{c_s}=1/c_T$
\cite{Creminelli:2014wna}\cite{Baumann:2015xxa}, the scalar power
spectrum is also actually the same as that in the disformal
frame, i.e., that of slow-roll inflation $P_\zeta={H_{inf}^2\over
8\pi^2M_p^2 \epsilon}$, as in \cite{Creminelli:2014wna}.

In Fig.\ref{fig04}, $\tilde H$ rapidly increases for a while,
similar to that in the so-called phantom inflation scenario
\cite{Piao:2003ty}. Thus, it is natural that the corresponding
$P_T$ is blue tilted. However, in the disformal frame, the
blue tilt is attributed to the rapid decrease of $c_T$. Thus, the
nontrivial effect of $c_T$ provides an alternative angle of
view for the phantom superinflation; i.e., it might be
just a slow-roll inflation living in a disformal frame
with $c_T$ rapidly decreasing. On the other hand, if we
treat it as $c_T=1$, what we will feel is just the
superinflation, and obviously there is not the ghost instability.

\section{Discussion}

We have illustrated that the effects of the propagation speed
$c_T$ of the primordial GWs on the inflationary GWs
spectrum may be nontrivial. We numerically confirmed that,
although $c_T(t)$ in the disformal frame may be set as
unity with a disformal redefinition of metric, the power spectrum
in the frame with $c_T=1$ is completely the same as that in
the original disformal frame, i.e., the oscillating shape in the
power spectrum is still reserved, since in the frame with $c_T=1$
the effect of $c_T$ is actually encoded in the nontrivially
varying Hubble parameter. Actually, the physical behavior of
horizon crossing, which depends on both the propagation speed
$c_T$ and the Hubble horizon, should be invariant in both frames.

However, if the condition $\tilde{\epsilon }\ll 1$ is imposed,
i.e., the Universe must be in slow-roll inflation in the frame
with $c_T=1$, the result will be the same as that in
Ref.\cite{Creminelli:2014wna}. Actually, this condition implicitly
assumes $\frac{\dot{c_T}}{Hc_T}\ll1$, and thus it is impossible that
the effect of the nontrivial variation of $c_T$ is involved. Here,
the key point is that the slow-roll inflation model with a
dramatic variation of $c_T(t)$ in the disformal frame might no
longer be the slow-roll inflation when we see it with
$c_T\equiv1$, which is different from the assumption made in
\cite{Creminelli:2014wna}. Namely, the result of \cite{Creminelli:2014wna} is
applicable only for the slow-roll inflation model where $c_T$ varies
very slow and the background is still slow-roll inflation
in the frame with $c_T=1$ after the disformal redefinition of the
metric. However, it is interesting to notice that for
(\ref{action}) the power spectrum of the scalar perturbation is not
affected by $c_T(t)$, so its spectrum is still the standard result
of slow-roll inflation no matter in which frame we see it, which
may be consistent with the observations.

Thus the nontrivial effect of $c_T$ may offer the slow-roll
inflation scenario more flexibility in predictions for the
primordial GWs. We present such a slow-roll model, in which the
scalar spectrum is scale invariant with slightly red tilt while
the local spectrum of GWs may be slightly or strongly blue, by
imposing a rapidly decreasing $c_T$. Although when we work in the
frame with $c_T=1$, what we will feel is the superinflation,
obviously there is no ghost instability.
The relevant issue
is interesting for further study.

The local oscillation in the power spectrum of primordial
GWs induced by $c_T(t)$ may be measurably imprinted in CMB B-mode
polarization spectrum\cite{Cai:2015ipa}\cite{Cai:2015dta}.
However, it is also possible that the induced oscillations
happened at a scale far smaller than the CMB scale.
Recently,  the LIGO Scientific Collaboration reported their direct detection of
gravitational waves which come from a binary black hole merger \cite{Abbott:2016blz}. There is no doubt that this discovery is of great significance in fundamental science and will be a scientific milestone for exploring our Universe.

\begin{figure}[htbp]
\includegraphics[scale=2,width=0.6\textwidth]{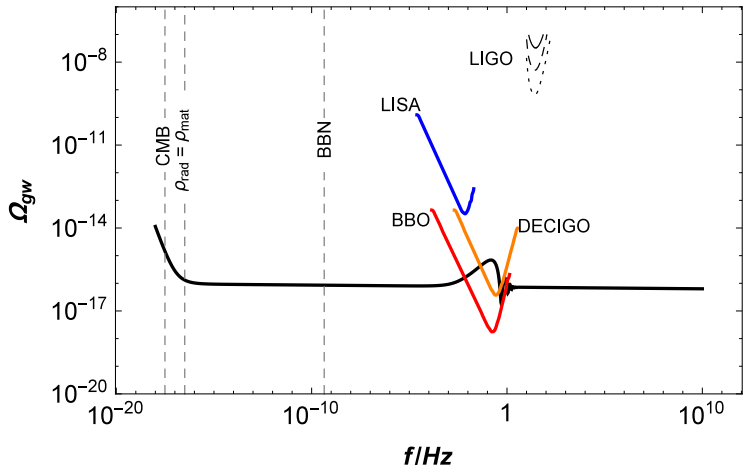}
\caption{$\Omega_{gw}$ with respect to frequency $f$. We take
$c_T(\alpha)=1-c_{T*}e^{-D (\alpha-\alpha_*)^2}$, where $D=0.2$
and $c_{T*}=0.9$.} \label{fig05}
\end{figure}

The local oscillation in the power spectrum or a blue power spectrum of primordial GWs will significantly boost the stochastic GWs background at the frequency band of Advanced
LIGO/Virgo (see \cite{Cai:2016ldn} for a recent study) as well as the space-based detectors such as BBO or
DECIGO.
We plot the energy spectrum $\Omega_{gw}(f)$ of the
primordial GWs in Fig.\ref{fig05}, where $f$ is the frequency of
GWs. In addition, it is also interesting to investigate the
effect of the corresponding oscillation on the circular
polarization of primordial GWs, e.g., \cite{Satoh:2007gn}.
Therefore, if these high-precision observations would observe the
oscillation shape or a blue tilt of the primordial GWs power spectrum, we will
promisingly acquire significant insights into the physics beyond
GR.

\textbf{Acknowledgments}

This work is supported by NSFC, Grant No.11222546, and National Basic
Research Program of China, Grant No.2010CB832804, and the Strategic
Priority Research Program of Chinese Academy of Sciences,
Grant No.XDA04000000.

\end{document}